%% file: main.tex
\documentclass[headings=standardclasses]{scrartcl}

\usepackage{geometry}
\usepackage[utf8]{inputenc}
\usepackage[T1]{fontenc}
\usepackage[onehalfspacing]{setspace}
\usepackage{times}
\usepackage[charter]{mathdesign}
\usepackage{microtype}
\usepackage{amsmath}
\usepackage{amsthm}
\usepackage{tabulary}
\usepackage{longtable}
\usepackage{booktabs} 
\usepackage{diagbox}
\usepackage{ragged2e}

\usepackage{siunitx}
\usepackage{bm}
\usepackage[pdftex]{graphicx}
\usepackage{rotating}
\usepackage{pdfpages}
\usepackage{pdflscape}
\usepackage{color}
\usepackage{array}
\usepackage[font=bf]{caption}
\usepackage{subcaption}
\usepackage{pgfplots}
\usepackage{pgfplotstable}
\usepackage{float}
\usepackage[title]{appendix}
\pgfplotsset{compat=1.9}
\usepackage{url}
\usepackage[section]{placeins}
\usepackage{natbib}
\usepackage{hyperref}
\hypersetup{
	colorlinks   = true,    
	urlcolor     = blue,    
	linkcolor    = blue,    
	citecolor    = blue      
}
\usepackage{pifont}
\newcommand{\cmark}{\ding{51}}
\newcommand{\xmark}{\ding{55}}

\makeatletter
\renewcommand*\env@matrix[1][c]{\hskip -\arraycolsep
  \let\@ifnextchar\new@ifnextchar
  \array{*\c@MaxMatrixCols #1}}
\makeatother

\usepackage{xcolor}
\usepackage{soul}

\usepackage{verbatim}

\begin{document}
\thispagestyle{empty}
\begin{spacing}{1.2}
\begin{flushleft}
\huge \textbf{A New Spatial Count Data Model with Bayesian Additive Regression Trees for Accident Hot Spot Identification} \\
\vspace{\baselineskip}
\normalsize
24 May 2020 \\
\vspace{\baselineskip}
\textsc{Rico Krueger}\textsuperscript{*} (corresponding author) \\
Transport and Mobility Laboratory \\
Ecole Polytechnique F\'{e}d\'{e}rale de Lausanne, Switzerland \\
rico.krueger@epfl.ch \\
\vspace{\baselineskip}
\textsc{Prateek Bansal}\textsuperscript{*} \\
Department of Civil and Environmental Engineering\\
Imperial College London, UK \\
prateek.bansal@imperial.ac.uk \\
\vspace{\baselineskip}
\textsc{Prasad Buddhavarapu} \\
Department of Civil Architectural and Environmental Engineering \\
The University of Texas at Austin, United States \\
prasad.buddhavarapu@utexas.edu \\
\vspace{\baselineskip}
\textsuperscript{*} Equal contribution.
\end{flushleft}
\end{spacing}

\newpage
\thispagestyle{empty}
\section*{Abstract}

The identification of accident hot spots is a central task of road safety management. Bayesian count data models have emerged as the workhorse method for producing probabilistic rankings of hazardous sites in road networks. Typically, these methods assume simple linear link function specifications, which, however, limit the predictive power of a model. Furthermore, extensive specification searches are precluded by complex model structures arising from the need to account for unobserved heterogeneity and spatial correlations. Modern machine learning (ML) methods offer ways to automate the specification of the link function. However, these methods do not capture estimation uncertainty, and it is also difficult to incorporate spatial correlations. In light of these gaps in the literature, this paper proposes a new spatial negative binomial model, which uses Bayesian additive regression trees to endogenously select the specification of the link function. Posterior inference in the proposed model is made feasible with the help of the P\'olya-Gamma data augmentation technique. We test the performance of this new model on a crash count data set from a metropolitan highway network. The empirical results show that the proposed model performs at least as well as a baseline spatial count data model with random parameters in terms of goodness of fit and site ranking ability.\\
\\
\emph{Keywords:} accident analysis, site ranking, spatial count data modelling, negative binomial model, Bayesian additive regression trees, P\'olya-Gamma data augmentation.

\newpage
\pagenumbering{arabic}

\section{Introduction}


The identification of accident-prone locations (so-called hot spots) is a core task of road safety management \citep{cheng2020exploring,huang2009empirical,lee2020optimal}. Of the various approaches for accident hot spot identification \citep[see][for a comparison]{wang2018evaluation, zahran2019spatial}, crash frequency analysis is the most widely employed method. Crash count data models are used to produce model-based rankings of hazardous sites and to predict crash counts at hot spots under counterfactual traffic flow and road design scenarios \citep{deacon1975identification}. Recent work also uses multivariate analysis for the joint modelling of road fatality and injury counts \citep{besharati2020bivariate}, single and multi-vehicle crashes \citep{wang2019freeway}, and crash frequency by travel modes \citep{huang2017multivariate}.    

Crash counts are typically modelled using Poisson log-normal or negative binomial regression models \citep{lord2010statistical}. Accommodating flexible representations of unobserved heterogeneity in model parameters and accounting for correlations between spatial units are central themes of the recent crash count modelling literature \citep{cai2019integrating,cheng2020exploring, dong2016macroscopic, heydari2016bayesian, mannering2016unobserved, ziakopoulos2020review}. However, these flexible representations of unobserved heterogeneity are achieved at the cost of a restrictive linear specification of the link function. Whilst linear-in-parameters link functions are appealing from an interpretability perspective, an over-simplification of the relationship between predictors and the explained variable may negatively affect the predictive performance of a model \citep{li2008predicting, huang2016predicting}.  

Since the predictive performance of a model is of paramount importance in hot spot identification and site ranking applications, the specification of the link function should be carefully selected. However, in practice, the space of possible link function specifications is prohibitively large, which precludes exhaustive specification searches. Modern machine learning (ML) methods offer a remedy to this challenge, as they enable automatic specification searches. A few studies have adopted kernel-based regression \citep{thakali2016nonparametric}, neural networks \citep{chang2005analysis, huang2016predicting, xie2007predicting, zeng2016modeling, zeng2016rule}, support vector machine \citep{dong2015support, li2008predicting}, and deep learning architectures \citep{cai2019applying, dong2018improved} for crash count modelling. 

Modern ML methods are shown to surpass the traditional count data models in terms of predictive accuracy but succumb to four limitations. First, with exception of the work by \cite{dong2015support}, none of the existing ML studies account for spatial correlations between observations. It is important to note that a non-linear link function specification does not inherently account for spatial correlations, and ignoring these correlations can deteriorate the robustness of predictions \citep{dong2015support}. Second, unlike traditional count data models, ML methods do not provide a quantification of estimation uncertainty and thus do not offer straightforward ways to construct confidence or credible intervals. Third, ML methods are fully nonparametric, with no easy ways to integrate interpretable components in link functions. In other words, if a user is interested in inferring the relationship between selected explanatory variables and crash counts, there is no straightforward way to specify a semiparametric link function in the above-mentioned ML-based count data models. Fourth, ML-based crash count studies benchmark the performance of their methods against simplistic parametric models which do not account for unobserved and spatial heterogeneity, which is not a fair comparison. More specifically, none of the previous studies address an important question---whether a count data model with a nonparametric link function can outperform a model with a linear link function that also accounts for unobserved heterogeneity. 

We emphasise that the before-mentioned ML methods adopt classical inference approaches, which yield point estimates of parameters of interest. On the contrary, the Bayesian approach facilitates accounting for various sources of uncertainty in model formulation and inference. For instance, posterior draws of site rankings at each iteration of the Gibbs sampler can directly provide ranking estimates with credible intervals \citep{miaou2005bayesian}. Hence, the fully Bayesian approach has emerged as the workhorse method in the site ranking literature.  

Along the same lines, this paper proposes a Bayesian negative binomial regression model, which not only addresses the first three limitations of the above-discussed ML methods by retaining all advantages of the statistical crash count models (interpretability, inference, accounting for random parameters and spatial correlations) but also allows for an additional nonparametric component in the link function with an endogenous selection of the specification. This nonparametric component is specified as sum-of-trees using the Bayesian Additive Regression Trees (BART) framework \citep{chipman2010bart}. The sum-of-trees specification inherently partitions the support of each explanatory variable during the estimation, resulting in a sum of step functions of individual predictors. Furthermore, if a tree depends only on one predictor, then this specification captures the main effect of the predictors; it represents an interaction effect, if a tree depends on more than one predictor. This process is equivalent to creating categorical variables from continuous variables, as it inherently accounts for interaction effects between predictors, while cut-offs and functional forms of interaction effects are endogenously selected during estimation based on predictive accuracy. This is particularly relevant in the context of the site ranking, where continuous predictors like speed limit and shoulder width are often converted into categorical variables by manually selecting cut-offs before entering into linear link functions because such explanatory variables are unlikely to have a constant marginal effect over the entire support. 

In the Gibbs sampler for the proposed model, we adopt the P{\'o}lya-Gamma augmentation method to deal with the non-conjugacy of the negative binomial likelihood \citep{polson2013bayesian}. The key idea of this data augmentation approach is to translate the negative binomial likelihood into a Gaussian likelihood by introducing auxiliary P{\'o}lya-Gamma-distributed random variables into the model. Finally, we address the last limitation of recent ML studies by providing a fair comparison of the proposed model with its linear-link counterpart, while also incorporating random parameters and spatial random effects. The proposed approach is a full Bayes (FB) method and the superiority of FB over empirical Bayes (EB) methods is well established in the literature \citep{guo2019comparative}. Nonetheless, we still benchmark our approach against EB in site ranking analysis.     

The remainder of this paper is organised as follows: In Section \ref{section:model_formulation}, we introduce the formulation of the proposed model framework and in Section \ref{section:estimation}, we outline the estimation approach. In Section \ref{section:empirical_analysis}, we present the empirical analysis and in Section \ref{section:conclusion}, we conclude and discuss avenues of future research.

\section{Model formulation} \label{section:model_formulation}

Let $y_{i}$ denote an observed crash count on road segment $i = \{1,2,\dots,N\}$. We consider the distribution of $y_{i}$ to be negative binomial (NB) with probability parameter $p_i$ and shape parameter $r$. The link function $\psi_{i} = \log \frac{p_{i}}{1-p_i}$ depends on road-segment-specific attributes $\bm{F}_{i}$ and $\bm{X}_{i}$. While $\bm{F}_{i}$ enters in the link function in a linear, interpretable form, the effect of $\bm{X}_{i}$ is specified using a sum of trees $\sum_{j=1}^{m}g_{i}(\bm{X}_{i}; \bm{T}_{j},\bm{M}_{j})$ \citep{chipman2010bart}. Here, $\bm{T}_{j}$ is a binary regression tree, and $\bm{M}_{j}$ denotes the parameters associated with its terminal nodes. To account for spatial correlations between road segments, we also include a spatial random effect $\phi_{i}$ into the link function. The collection of spatial random effects $\bm{\phi} = \left ( \phi_{1}, \ldots \phi_{N} \right)^{\top}$ follows a matrix exponential spatial specification (MESS) of dependence that ensures exponential decay of influence over space \citep{lesage2007matrix}. Consequently, we have $\bm{S} \bm{\phi} = \exp(\tau \bm{W})\bm{\phi} = \bm{\epsilon}$ with $ \bm{\epsilon} \sim \text{Normal}(0,\sigma^2 \bm{I}_{N})$, where $\bm{W}$ is an $N \times N$ non-negative spatial weight matrix, $\tau$ captures the magnitude of spatial association, and $\sigma$ is an error scale.

The proposed model is summarised below:  
\begin{align}
& y_{i} \sim \text{NB}(r,p_i), & & i = 1,\dots,N \\
& p_i = \frac{\exp(\psi_i)}{1+\exp(\psi_i)}, & & i = 1,\dots,N\\
& \psi_i = \bm{F}_{i}^{\top} \bm{\gamma} +  G_{i}(\bm{X}_{i}; \bm{T},\bm{M})  + \phi_{i},  & & i = 1,\dots,N   \label{eq_psi}\\
& G_{i}(\bm{X}_{i}; \bm{T},\bm{M}) = \sum_{j=1}^{m}g_{i}(\bm{X}_{i}; \bm{T}_{j},\bm{M}_{j}),  & & i = 1,\dots,N  \label{eq_tree}\\
& \bm{S}\bm{\phi} = \exp(\tau \bm{W})\bm{\phi} =  \bm{\epsilon} \\
& \bm{\epsilon} \sim \text{Normal}(0,\sigma^2 \bm{I}_{N}) 
\end{align} 
Equation \ref{eq_psi} can be rewritten in vector form. We have
\begin{equation}
\bm{\psi} = \bm{F}\bm{\gamma} +  \bm{G}(\bm{X}; \bm{T},\bm{M}) + \bm{\phi}
\end{equation}
such that
\begin{equation}
P(\bm{\psi} \lvert \bm{\gamma} , \bm{T}, \bm{M},\sigma^{2}, \tau)  = (2\pi\sigma^2)^{-\frac{N}{2}} \exp\left( -\frac{ [\bm{\psi} - \bm{F}\bm{\gamma} - \bm{G}]^{\top} \Tilde{\bm{\Omega}} [\bm{\psi} - \bm{F}\bm{\gamma} - \bm{G}]}{2} \right)
\end{equation}
with $\Tilde{\bm{\Omega}} = \frac{\bm{S}^{\top} \bm{S}}{\sigma^2}$.

\subsection{Bayesian additive regression trees (BART)}

\cite{chipman2010bart} introduce BART as a nonparametric prior over a regression function to capture non-linear relationships and interaction effects between predictors. BART has been widely applied to a wide range of regression and classification problems \citep[see][]{hill2020bayesian} but the current study presents the first application of BART in spatial count data modelling.  

\begin{figure}[H]
\centering
\includegraphics[width = 0.8 \textwidth]{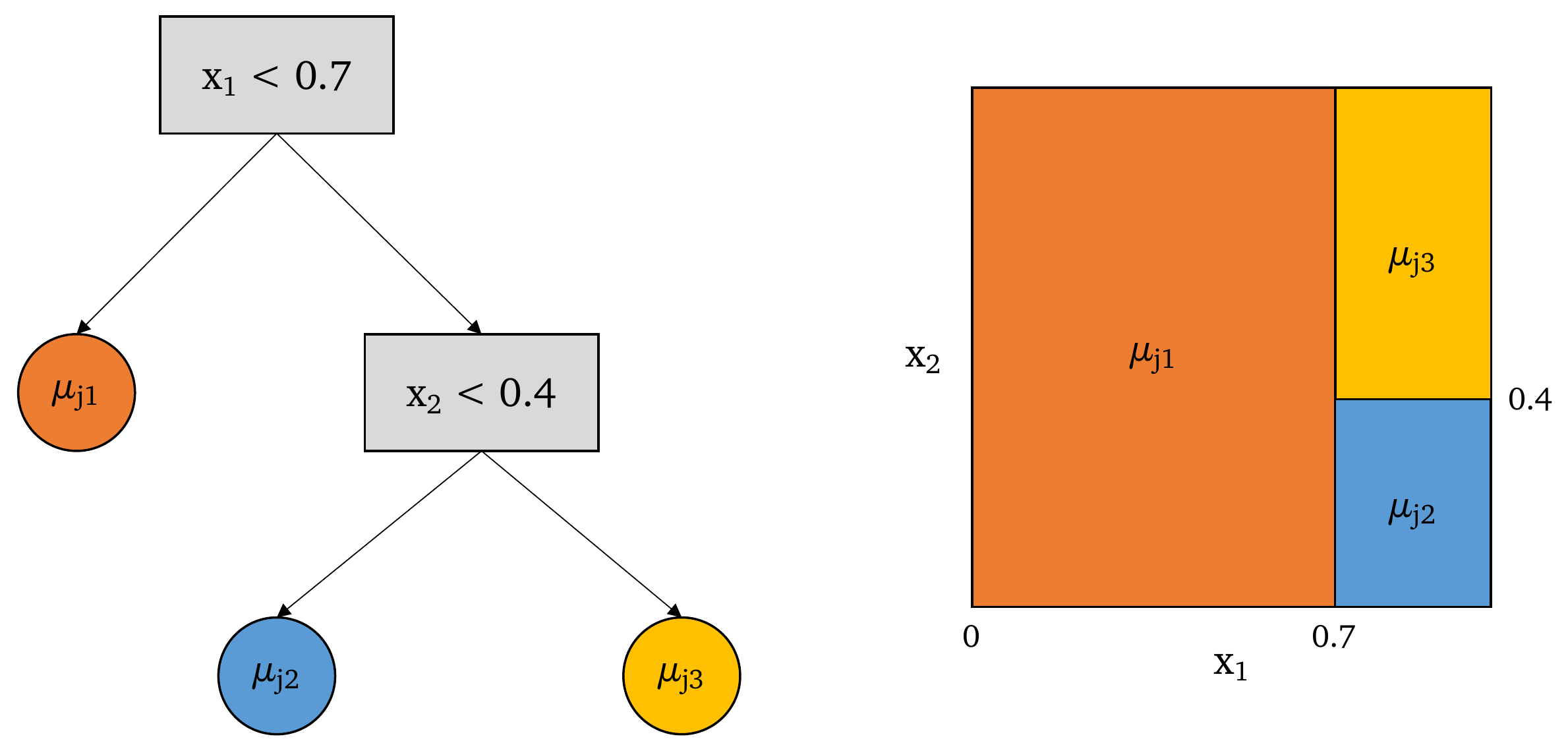}
\caption{Illustration of a single binary tree $\bm{T}_{j}$ and its corresponding partition of the predictor space $\bm{X} = \{ x_{1}, x_{2} \}$. Internal tree nodes are marked with their splitting rules; leaf nodes are marked with their leaf parameters.} \label{fig:tree}
\end{figure}

BART specifies the regression function as a sum of trees. Each tree $\bm{T}_{j}$ consists a group of decision nodes with splitting rules and a set of terminal leaf nodes. Figure \ref{fig:tree} illustrates one such tree, whose splitting rules at the decision nodes are $x_{1}<0.7$ and $x_{2}<0.4$, and whose leaf nodes are $\bm{M}_{j} = \{\mu_{j1}, \mu_{j2}, \mu_{j3}\}$. A tree and its associated decision rules create partitions of the predictor space such that each partition corresponds to a leaf node. In the illustrative example, three partitions are induced: $\mathcal{P}_{j1} = \{x_{1} < 0.7\},\mathcal{P}_{j2} = \{x_{1}>0.7, x_{2}<0.4\},\mathcal{P}_{j3} =  \{x_{1}>0.7, x_{2}>0.4\}$. The illustrative tree inherently accounts for the interaction between $x_{1}$ and $x_{2}$. However, if a tree depends only on one predictor, the main effect of that predictor is captured. The function for the illustrative tree is:
\begin{equation}
g(\bm{X}; \bm{T}_{j},\bm{M}_{j}) = \mu_{jt} \;\;\; \text{if} \;\;\; \bm{X} = \{x_{1},x_{2}\} \in \mathcal{P}_{jt} \;\;\; \forall t \in \{1,2,3\}.
\end{equation}
$g(\bm{X}; \bm{T}_{j},\bm{M}_{j})$ is a step function of a subset of predictors and thus, BART is a sum of step functions. The key idea of BART is to regularise the fit by keeping the effect of each individual tree small such that each of tree can explain a different portion of the regression function. In other words, each tree is constrained to be a weak learner and BART is an ensemble of weak learners. The sum of trees model is a flexible specification, which results in an excellent predictive accuracy \citep{chipman2010bart}. Furthermore, splitting rules and leaf nodes are estimable parameters in BART, which in turn facilitates automated specification searches through endogenous partitioning of the predictor space. 
 
\subsection{Spatial error dependence}

In this study, we adopt the matrix exponential spatial specification \citep[MESS;][]{lesage2007matrix} to model the unobserved dependence among spatial units. MESS is an appealing specification because it offers computational advantages by simplifying the log-likelihood computations \citep{lesage2009introduction}. MESS also has close correspondence with spatial autoregressive (SAR) and conditional autoregressive (CAR) specifications \citep{strauss2017matrix}. Whereas SAR and CAR assume geometric decay of spatial correlation, MESS relies on exponential decay \citep{lesage2007matrix}. Readers can refer to \cite{debarsy2015large} for a detailed derivation of the large sample properties of MESS. 

\section{Estimation} \label{section:estimation} 

\subsection{P{\'o}lya-Gamma data augmentation}

Irrespective of the link function specification, conditional posterior distributions of the parameters of the negative binomial model do not belong to a known family of distributions \citep{buddhavarapu2016modeling,park2009application}. In such situations, data augmentation techniques, which involve intermediate or additional latent random variables, are used to derive tractable conditional posteriors of model parameters \citep{van2001art}. BART was originally developed for Gaussian models and therefore, its application in non-Gaussian models can be operationalised by transforming the original model into a Gaussian model via data augmentation. For example, \cite{chipman2010bart} extend BART for a binary probit model with a flexible link function and estimated it by adopting the data augmentation technique suggested by \cite{albert1993bayesian}. Similarly, \cite{kindo2016multinomial} use BART to specify indirect utilities in a multinomial probit model and estimated it by augmenting indirect utilities as latent Gaussian random variables.  

In this study, we adopt the P{\'o}lya-Gamma data augmentation technique which transforms the negative binomial model into a heteroskedastic Gaussian model \citep{polson2013bayesian}. This augmentation is not only appropriate for any negative binomial specification, as shown by \citet{buddhavarapu2016modeling}, but also enables seamless integration of the Gibbs sampler of the heteroskedastic BART \citep{bleich2014bayesian} into that of the proposed model. 

P{\'o}lya-Gamma data augmentation introduces auxiliary P{\'o}lya-Gamma random variates $\omega_{i}$ into the model. Conditional on $\omega_{i}$, the negative binomial likelihood is translated into a heteroskedastic Gaussian likelihood, while closed-form posteriors for the augmentation variables are retained \citep[see][for a detailed proof]{polson2013bayesian}.   
\begin{align}
        \omega_{i} &\sim \text{PG}(y_{i} + r,0) \\
        P(y_{i} \lvert \psi_{i},r,\omega_{i}) & \propto \exp\left(- \frac{\omega_{i}}{2}\left[\psi_{i} - \frac{y_{i}-r}{2\omega_{i}} \right]^2\right) \\
        P(\bm{y} \lvert \bm{\psi},r,\bm{\omega}) & \propto \exp\left( -\frac{1}{2}[\bm{\psi} - \bm{Z}]^{\top} \bm{\Omega} [\bm{\psi} - \bm{Z}]\right) 
\end{align}
where 
\begin{equation}
 \bm{Z} = \begin{bmatrix} 
    \frac{y_{1}-r}{2\omega_{1}}\\
   \vdots\\
    \frac{y_{N}-r}{2\omega_{N}}\\
    \end{bmatrix}_{N \times 1}
\bm{\Omega} = \begin{bmatrix} 
   \omega_{1} & \dots & 0 \\
    \vdots & \ddots &  \vdots\\
    0 &    \dots    &  \omega_{N} 
    \end{bmatrix}_{N \times N}
\end{equation}
\begin{equation}\label{eq:aug}
\bm{Z} = \bm{\psi} + \bm{\varrho} =  \bm{F}\bm{\gamma} + \bm{G}(\bm{X}; \bm{T},\bm{M}) + \bm{\phi} + \bm{\varrho}, \quad \bm{\varrho} \sim \text{Normal}(\bm{0},\bm{\Omega}^{-1}) 
\end{equation}

\subsection{Prior specification}

We adopt the strategy used by \cite{chipman2010bart} to specify priors on $\bm{T}_{j}$ and $\bm{M}_{j} \lvert \bm{T}_{j}$. Other prior distributions are summarised below: 
\begin{align}
& \bm{\gamma} \sim \text{Normal}(\bm{\zeta}_{\bm{\gamma}}, \bm{\Delta}_{\bm{\gamma}}) \\
& \tau \sim \text{Normal}(\zeta_{\tau}, \sigma_{\tau}^2) \\
& \sigma^{-2} \sim \text{Gamma} (b_{\sigma^2},c_{\sigma^2}) \\
& r \sim \text{Gamma}(r_{0},h) \\
& h \sim \text{Gamma}(b_{0},c_{0}) 
\end{align} 
Here $\{\bm{\zeta}_{\bm{\gamma}}, \bm{\Delta}_{\bm{\gamma}}, \zeta_{\tau}, \sigma_{\tau}^2,  b_{\sigma^2},c_{\sigma^2}, r_{0}, b_{0},c_{0}\} \; \cup \; \{\text{Tree Hyper-parameters}\}$ is a set of hyper-parameters and $\bm{\Theta} = \left\{ \bm{\gamma}, \bm{\phi}, \bm{T}, \bm{M}, \sigma^{2}, \bm{\omega}, r, h, \tau \right\}$ is a set of latent variables of the models. Thus, the joint distribution of observed and latent variables is: 
\begin{equation}
    \begin{split}
        P(\bm{y} , \bm{\Theta}) & =  P(\bm{y} \lvert r,\bm{\omega}, \bm{\gamma}, \bm{T}, \bm{M}, \bm{\phi}) P(\bm{\phi} \lvert \sigma^{2}, \tau)  P(r \lvert r_{0},h) P(h\lvert b_{0},c_{0}) P(\sigma^{-2}  \lvert b_{\sigma^2},c_{\sigma^2}) P(\tau \lvert \zeta_{\tau}, \sigma_{\tau}^2)\dots \\
        & \dots  P(\bm{\gamma} \lvert \bm{\zeta}_{\bm{\gamma}}, \bm{\Delta}_{\bm{\gamma}})  P(\bm{T}, \bm{M} \lvert \{\text{Tree Hyper-parameters})  \left(\prod_{i=1}^N P(\omega_{i}\lvert r)\right)
    \end{split}
\end{equation}

\subsection{Posterior updates}

To infer the posterior distributions of the model parameters of interest, we construct Markov chains by generating samples from the conditional distributions of individual coordinates of the parameters space. One iteration of the Gibbs sampler is described below:
\begin{itemize}
    \item Update $\bm{\phi}$ by sampling $\bm{\phi} \sim \text{Normal}\left((\bm{\Omega} + \Tilde{\bm{\Omega}})^{-1}\bm{\Omega}(\bm{Z} - \bm{G} - \bm{F}\bm{\gamma}), (\bm{\Omega} + \Tilde{\bm{\Omega}})^{-1}\right)$. 
    \item Update $\bm{\gamma}$ by sampling from \\
    $\bm{\gamma} \sim \text{Normal}\left( (\bm{\Delta}_{\bm{\gamma}}^{-1} + \bm{F}^{\top}\bm{\Omega}\bm{F})^{-1}(\bm{F}^{\top} \bm{\Omega} (\bm{Z} - \bm{G}  - \bm{\phi}) + \bm{\Delta}_{\bm{\gamma}}^{-1}\bm{\zeta}_{\bm{\gamma}}), (\bm{\Delta}_{\bm{\gamma}}^{-1} + \bm{F}^{\top}\bm{\Omega}\bm{F})^{-1}\right)$
    \item Update $\sigma^{-2}$ by sampling $\sigma^{-2} \sim \text{Gamma}\left(b_{\sigma^2} + \frac{N}{2},  c_{\sigma^2} + \frac{ \bm{\phi}^{\top} \bm{S}^{\top} \bm{S} \bm{\phi}}{2} \right)$. 
    \item Update $\omega_{i}$ by sampling from $\omega_{i} \sim \text{PG}(y_{i} + r, \psi_i)$. 
    \item Update $r$ by sampling from a Gamma distribution \citep[see][Section 4.1.1]{zhou2012lognormal}.   
    \item Update $h$ by sampling from $ h \sim  \text{Gamma}(r_{0} + b_{0},  r + c_{0})$.
    \item Update $\tau$ using the Metropolis–Hastings algorithm where $ P(\tau  \lvert \cdot) \propto \exp \left ( - \frac{(\tau - \zeta_{\tau})^{2}}{2 \sigma_{\tau}^2} \right )\exp\left( -\frac{ \bm{\phi}^{\top} \Tilde{\bm{\Omega}} \bm{\phi}}{2} \right)$.
    \item Update $\bm{T}$ and $\bm{M}$ as illustrated for the heteroskedastic BART by \cite{bleich2014bayesian}, where the dependent variable is $\bm{Z} - \bm{\phi} - \bm{F}\bm{\gamma}$ and the error covariance matrix is $\bm{\Omega}^{-1}$.  The updates of BART parameters are based on an iterative Bayesian backfitting algorithm \citep{hastie2000bayesian}, where one tree is updated conditional on all other trees. 
    \item Compute $Z_{i} = \frac{y_{i}-r}{2\omega_{i}}, \quad i = {1, \dots, N}$.
\end{itemize}

In this study, we present the estimation procedure for a specification with only non-random parameters in the interpretable part of the link function. However, a modeller might expect unobserved heterogeneity in some elements of $\bm{\gamma}$. For completeness, we note that the proposed estimation procedure can be extended for such specifications of unobserved heterogeneity to facilitate flexibility at no compromise in interpretability. Thanks to P{\'o}lya-Gamma data augmentation, the extension is as simple as deriving a Gibbs sampler for a linear mixed-effects model. 

\section{Empirical analysis} \label{section:empirical_analysis}

\subsection{Data}

The proposed model framework is empirically validated using detailed crash frequency data from 1,158 contiguous road segments of eleven highway facilities in the Greater Houston metropolitan area in the United States of America.
The data were compiled by geographically fusing information retrieved from accident and road databases for an observation period covering four consecutive calendar years in the period from 2007 to 2010. 
For each road segment, the considered data include the annual crash count aggregated over all accident types and severity levels as well as other segment-specific characteristics, namely the type of highway facility the segment in question is associated with, the traffic volume, the truck traffic percentage, the road condition and roadway design attributes. The identified 
Table \ref{tab:sample_description} enumerates summary statistics of the annual crash counts and the segment features for the considered road segments for each year of the observation period. 
More details about the the data collection and processing are presented in \citet{buddhavarapu2015bayesian}.
In the subsequent analysis, we treat the data from each year as a separate sample and evaluate the empirical performance of the proposed model across the different years of the observation period.

\begin{landscape}
\begin{table}[H]
\centering
\footnotesize
\input{table_sample_description}
\caption{Sample description (N = 1,158)}
\label{tab:sample_description}
\end{table}
\end{landscape}

\subsection{Methodology}

\subsubsection{Model specifications}

We contrast the performance of four different model specifications. The proposed negative binomial Bayesian additive regression trees (\emph{NB-BART-I}, henceforth) model is benchmarked against negative binomial regression models with spatial error terms and linear-in-parameters link functions. We consider one model with fixed parameters (\emph{NB-fixed}, henceforth) as well as a model whose linear-in-parameters link function contains both fixed and random parameters (\emph{NB-random}, henceforth). The specifications \emph{NB-BART-I}, \emph{NB-fixed} and \emph{NB-random} use a restricted predictor space, as indicated in column ``Predictor space: I'' of Table \ref{tab:sample_description}. In the restricted predictor space, some continuous predictors are converted into dummy variables, because assuming a constant marginal effect over the entire support of the predictors is not meaningful. For example, left shoulder width is a continuous variable but its support is partitioned at 10 feet. Column ``Predictor space: I (random)'' in the same table indicates the predictors that are associated with fully-correlated random parameters in the link function. We conducted an extensive specification search, during which we closely monitored model tractability, to determine which predictors to associate with random parameters. We also test another specification of the proposed model (\emph{NB-BART-II}, henceforth), which considers all predictors in their original form, and allow BART to automatically partition the predictor space. Column ``Predictor space: II'' of Table \ref{tab:sample_description} indicates which predictors are included in specification \emph{NB-BART-II}. 

For the definition of the spatial weight matrix $\bm{W}$, we exploit the inherent neighbourhood structure of the road segments. First, we create a $N \times N$ contiguity-based proximity matrix $\bm{C}$, which is informed by the neighbourhood relationships of the road segments. An element $C_{ij}$ of $\bm{C}$ is given by
\begin{equation}
C_{ij} = 
\begin{cases}
\frac{1}{d_{i}(j)} & \text{for } d_{i}(j) \in \{1, \ldots, k^{*} \} \\
0 & \text{otherwise}
\end{cases},
\end{equation}
where $d_{i}(j)$ takes a value of $k$ if $i$ and $j$ are $k$th-order neighbours and zero otherwise. 
$k^{*} \in \left \{ x \in \mathbb{N} \vert x \in [1, N] \right \}$ is a truncation point, which is set to 3 in the current application. The intuition underlying the definition of $\boldsymbol{C}$ is that distant road segments exhibit weaker spatial correlation than proximal ones. Ultimately, we obtain $\bm{W}$ by row-normalising $\bm{C}$, i.e. $W_{ij} =\frac{C_{ij}}{\sum_{j} C_{ij}}$. 

\subsubsection{Implementation and estimation practicalities} 

We implement the MCMC algorithm outlined in Section \ref{section:estimation} by writing our own Python code.\footnote{The Python code is publicly available at \url{https://github.com/RicoKrueger/nb_bart}.} Our implementation of the heteroskedastic BART updates follows the documentation provided by \citet{bleich2014bayesian} and \citet{kapelner2016bart}. For the generation of the P\'olya-Gamma random variates, we rely on an existing Python implementation \citep{linderman2015dependent, linderman2016bayesian, linderman2016recurrent} of the appropriate sampling methods \citep{polson2013bayesian, windle2014sampling}.

For each of the considered model specifications, the MCMC algorithm is executed with four parallel Markov chains and 10,000 draws for each chain, whereby the initial 5,000 draws of each chain are discarded for burn-in. After the burn-in period, a thinning factor of 2 is applied. The step size of the Metropolis-Hastings algorithm for the generation of the posterior draws of the spatial association parameter $\tau$ is adaptively tuned with a target acceptance rate of 44\%, which is the recommended acceptance ratio for a one-dimensional target density \citep[see][]{roberts1997weak}. Convergence of the MCMC simulation is monitored with help of the potential scale reduction factor \citep{gelman1992inference}.

\subsubsection{Assessment of model fit} \label{subsection:assessment_fit}

We evaluate the goodness of fit of the considered methods using the log pointwise predictive density \citep[LPPD;][]{gelman2014understanding} and the root mean square error (RMSE):
\begin{itemize}
\item LPPD is a strictly proper scoring rule \citep{gneiting2007strictly}, which corresponds to the logarithm of the pointwise likelihood integrated over the posterior distribution of the relevant model parameters. It is given by 
\begin{equation}
\text{LPPD} = \sum_{i = 1}^{N}
\log \left (
\int P(y_{i} \vert \boldsymbol{\theta}_{i}) p(\boldsymbol{\theta}_{i} \vert \boldsymbol{y}) d \boldsymbol{\theta}_{i}
\right ),
\end{equation}
where $\boldsymbol{\theta}_{i} = \{ \psi_{i}, r \}$. A larger value of LPPD indicates superior goodness of fit.
\item RMSE is informed by the discrepancy between the predicted accident count $\hat{y}_{i}$ and the observed accident count $y_{i}$, whereby the former is given by the posterior mean of the conditional expectation of the crash count at site $i$. It is given by 
\begin{equation}
\text{RMSE} = \sqrt{\frac{1}{N} \sum_{i=1}^{N} \left( \hat{y}_{i} - y_{i} \right )^2}.
\end{equation}
\end{itemize}

\subsubsection{Assessment of site ranking ability} \label{subsection:assessment_ranking}

Numerous techniques for the model-based identification of accident hot spots are presented in the literature  \citep[see e.g.][]{aguero2009bayesian, geedipally2010identifying, miaou2005bayesian, miranda2007bayesian, washington2014applying}. In this study, we use the probabilistic ranking method proposed by \citet{schmidt2012modeling} to identify hazardous sites. To be specific, we rank sites by their posterior mean probability to belong to the top 5\% most hazardous sites in the network. Formally, we proceed as follows: In each MCMC iteration $d$, we calculate the probability that a site belongs to the top $m$ most hazardous sites, i.e. $P_{d} \left ( R(\lambda_{i}) \leq m \right )$, where $R(\lambda_{i})$ denotes the rank of site $i$ as a function of the expected accident count $\lambda_{i}$. The posterior mean of the crash statistic $\mathbb{E} \left ( R_{i} \leq m \vert \boldsymbol{y}, \boldsymbol{X} \right )$ is then obtained by averaging the respective posterior draws, and in the current application, we use $m = \alpha \cdot N$ with $\alpha = 5\%$. \citeauthor{schmidt2012modeling}'s (\citeyear{schmidt2012modeling}) probabilistic ranking method robustifies the Bayesian ranking procedures introduced by \citet{miaou2005bayesian} against heterogeneity in the posterior variance of the underlying decision parameter. 

Different hot spot identification methods can be compared by assessing the temporal consistency of the produced site rankings \citep{cheng2008new, montella2010comparative}. In this study, we employ the site consistency, method consistency and total rank differences tests of \citet{cheng2008new} to evaluate site ranking consistency
We use $H_{\alpha, t}$ to denote the set of sites that are identified as hazardous in time period $t$ at risk level $\alpha$ and define the site ranking consistency tests as follows:
\begin{itemize}
\item The site consistency test ($\mathcal{T}_{\text{SC}}$) measures the ability of a method to consistently identify a site as high risk over two consecutive time periods. It is based on the assumption that an unsafe site should remain hazardous provided that no safety improvements are implemented. $\mathcal{T}_{\text{SC}}$ for period $t$ is obtained by averaging the predicted accident counts $\hat{y}_{h, t+1}$ for period $t+1$ of all sites $h \in H_{\alpha, t}$:
\begin{equation}
\mathcal{T}_{\text{SC}, t} = \frac{1}{\vert H_{\alpha, t} \vert} \sum_{h \in H_{\alpha, t}} \hat{y}_{h, t+1}.
\end{equation}
A larger value of $\mathcal{T}_{\text{SC}, t}$ indicates superior site ranking consistency.
\item The method consistency test ($\mathcal{T}_{\text{MC}}$) measures the ability of a method to consistently identify the same hazardous sites over two consecutive time periods. $\mathcal{T}_{\text{MC}}$ for period $t$ is given by the cardinality of the intersection of $H_{\alpha, t}$ and $H_{\alpha, t+1}$, i.e.
\begin{equation}
\mathcal{T}_{\text{MC}, t} = \left \vert H_{\alpha, t} \cap H_{\alpha, t+1} \right \vert.
\end{equation}
A larger value of $\mathcal{T}_{\text{MC}, t}$ indicates superior site ranking consistency.
\item Finally, the total rank differences test ($\mathcal{T}_{\text{TRD}}$) measures the average absolute differences in ranks of hazardous sites over two consecutive time periods. $\mathcal{T}_{\text{TRD}}$ for period $t$ is given by
\begin{equation}
\mathcal{T}_{\text{TRD}, t} = \frac{1}{\vert H_{\alpha, t} \vert} \sum_{h \in H_{\alpha, t}} \left \vert R_{h, t+1} - R_{h, t} \right \vert,
\end{equation}
where $R_{h, t}$ denotes the rank of site $h$ in period $t$. 
A smaller value of $\mathcal{T}_{\text{TRD}, t}$ indicates superior site ranking consistency.
\end{itemize}

\subsection{Results}

\subsubsection{Model fit}

Table \ref{tab:fit} gives the goodness of fit measures of the considered model specifications across the four years of the observation period. It can be seen that NB-BART-II with an unrestricted predictor space outperforms the competing model specifications by a significant margin in all four samples. For example, for year 2010, NB-BART-II yields LPPD and RMSE values of $-3510.43$ and $10.53$, respectively, whereas the next-best model specification NB-BART-I yields LPPD and RMSE values of $-3558.87$ and $11.46$, respectively. Furthermore, we observe that NB-BART-I closely matches the performance of NB-random in the considered samples. As expected, NB-fixed exhibits the poorest performance among the four considered model specifications in all four samples due to its rigid link function specification.

In sum, Table \ref{tab:fit} shows that the endogenous partitioning of the predictor space induced by the BART component in NB-BART-II results in a substantial improvement in goodness of fit. We acknowledge that a modeller could exogenously partition the support of the continuous predictors in manifold ways to eventually achieve a comparable or better goodness-of-fit. However, the computation time of such an iterative hit-and-trial specification search would be prohibitive, as the space of possible specifications is enormously large. NB-BART-II automates this search process and achieves the optimal partitioning of the predictor space and thus obviates the need for expensive specification searches. Consequently, NB-BART-II with an unrestricted predictor space should be preferred over NB-BART-I with a restricted predictor space.

\begin{table}[H]
\centering
\small
\input{table_fit.tex}
\caption{Goodness of fit.} \label{tab:fit}
\end{table}

\subsubsection{Site ranking performance}

\begin{figure}[H]
\centering
\includegraphics[width = 0.85 \textwidth]{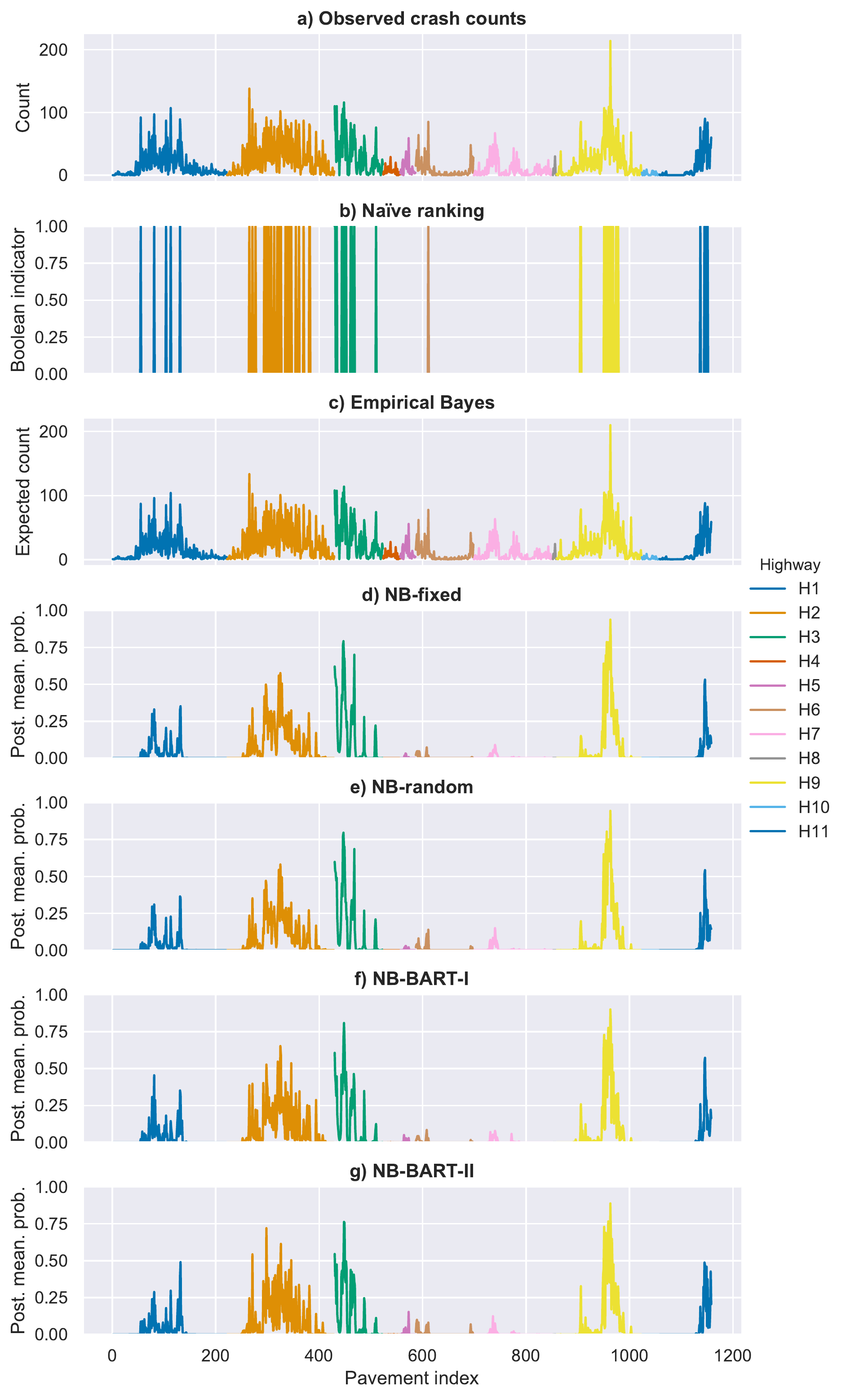}
\caption{Site ranking results for year 2010. The na\"ive ranking approach consists of identifying the top 5\% most hazardous sites based on the observed accident counts. For the probabilistic methods NB-fixed, NB-random, NB-BART-I and NB-BART-II, the plot shows the posterior mean probability that a site belongs to the top 5\% most hazardous sites. The highways are given arbitrary names for proprietary reasons.} \label{fig:hot_spots}
\end{figure}

Next, we compare the site ranking ability of the different model specifications. Our analysis covers all four years of the observation period but to avoid clutter, we restrict the visual display of the site ranking results to year 2010. 
Subplot a) of Figure \ref{fig:hot_spots} shows the observed crash count for each pavement segment. 
Subplot b) of the same figure displays the results of the na\"ive ranking approach, which involves assigning a value of 1 to the top 5\% sites with the highest accident count and a value of 0 to all other sites. 
Subplot c) shows the expected accident counts calculated via the empirical Bayes (EB) approach \citep{hauer2002estimating} in conjunction with a negative binomial model and maximum likelihood estimation.
Subplots d) to g) correspond to the four NB-BART specifications; each of the plots shows the posterior mean probabilities of each road segment to belong to the top 5\% most hazardous sites in the network. 

Figure \ref{fig:hot_spots} illustrates that the probabilistic ranking approaches can better capture the inherent uncertainty about the safety of a site, as the results differ markedly from the observed accident counts. In contradistinction, the results of the na\"ive approach and the EB method closely mimic the observed accident counts and thus provided limited additional information. Furthermore, it can be seen that the posterior mean probability plots corresponding to the specifications NB-fixed, NB-random, NB-BART-I, NB-BART-II are virtually indistinguishable from each other. This insight corroborates our earlier finding that the automated selection of an optimal link function specification in BART obviates the need for random parameters in a link function. Notwithstanding significant differences in goodness-of-fit (see Table \ref{tab:fit}), Figure \ref{fig:hot_spots} suggests that model specifications NB-BART-I and NB-BART-II compare favourably to each other as well as to NB-fixed and NB-random in terms of their site ranking abilities. 

Next, we contrast the site ranking ability of the EB approach as well as of NB-fixed, NB-random, NB-BART-I and NB-BART-II using the site consistency ($\mathcal{T}_{\text{SC}}$), method consistency ($\mathcal{T}_{\text{MC}}$) and total rank differences ($\mathcal{T}_{\text{TRD}}$) tests of \citet{cheng2008new}. Table \ref{tab:ranking} shows the results of the site ranking consistency tests for three reference period. We observe that NB-BART-II, followed by NB-BART-I, performs best in respect to site consistency for all reference periods. In terms of method consistency and total rank differences, none of the considered methods display superior performance across the three reference periods and the relative difference in the test score of the methods are not substantial. Therefore, we conclude that NB-BART affords similar site ranking consistency as NB-fixed and NB-BART. In addition, we note the EB approach underperforms in terms of site consistency but remains competitive with respect to the other methods in terms of method consistency and total rank differences. 

\begin{table}[H]
\centering
\small
\input{table_ranking_performance}

\caption{Site ranking consistency.} \label{tab:ranking}
\end{table}

\subsubsection{Variable importance and parameter estimates}

\begin{figure}[H]
\centering
\includegraphics[width = \textwidth]{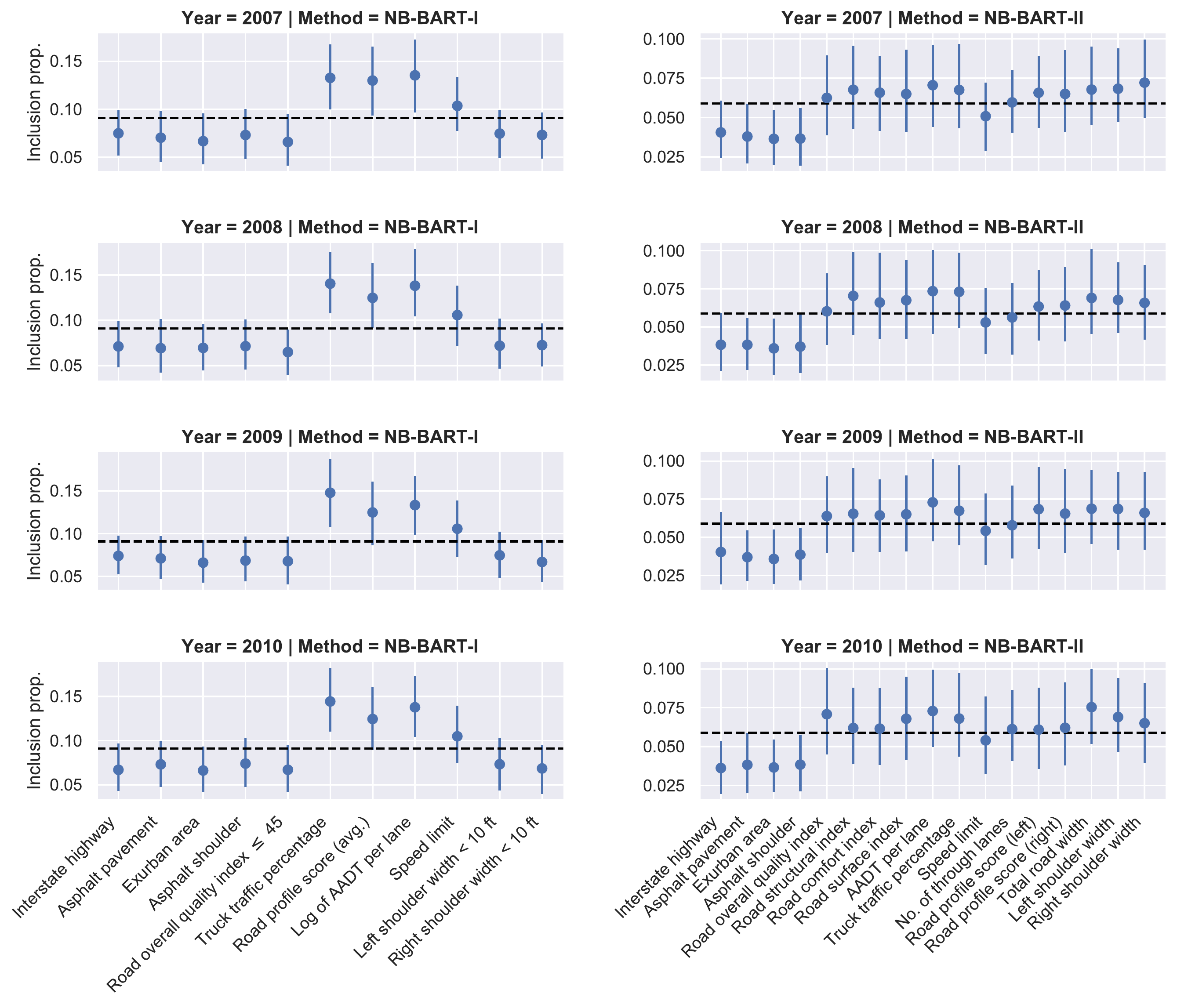}
\caption{Variable inclusion proportions. The dots mark the posterior means; the vertical error bars mark the 95\% credible intervals; the dashed horizontal lines indicate the equal importance inclusion proportions.} \label{fig:variable_inclusion}
\end{figure}

As a byproduct of the estimation, BART provides a statistic about the variable importance \citep[see][]{bleich2014variable}. In the context of BART, the notion of ``importance'' is different from elasticity. To be precise, variable importance of BART predictors denotes the proportion of times a variable of the predictor space is included in a splitting rule. Figure \ref{fig:variable_inclusion} shows the inclusion proportions of the predictors used in the model specifications NB-BART-I and NB-BART-II for the whole observation period. For both NB-BART specifications, we observe that the importance of the different variables is stable across the different years of the observation period. This insight shows that BART can consistently identify the variation explained by each predictor. Furthermore, the importance of several predictors (e.g. road quality index and left shoulder width) is higher in NB-BART-II than in NB-BART-I. This is because these variables are included as dummy variables in NB-BART-I, and variables with binary support only admit at most one split. In contrast, these variables are included as continuous predictors in NB-BART-II, and variables with continuous support inherently provide more ways to be incorporated into decision rules. This analysis suggests that the exogenous conversion of a continuous predictor into a dummy variable may reduce its importance. Yet, the endogenous partitioning of the predictor space in NB-BART-II allows to bypass the shortcomings of a manual specification selection.

Furthermore, Figure \ref{fig:variable_inclusion} shows the estimated posterior distribution of the negative binomial shape parameter $r$ (whose inverse is also referred to as the dispersion parameter) and the spatial association parameter $\tau$ for year 2010 of the observation period for each of the considered model specifications. For both model parameters, it can be seen that the posterior distributions produced by the different model specifications closely overlap. For the spatial association parameter $\tau$, we observe that the posterior distributions produced by NB-BART-I and -II are slightly shifted to the right relative the posterior distributions produced by the other model specifications. A possible explanation for this right shift is that the BART component absorbs some of the spatial dependence that is left unexplained by the other model specifications.

Finally, Table \ref{tab:results} displays the parameters estimates for NB-fixed and NB-random for all four years of the observation period. Along with the variable inclusion proportions shown in Figure \ref{fig:variable_inclusion}, Table \ref{tab:results} suggests that the importance of the predictors is stable across time periods. Consequently, temporal instability \citep{mannering2018temporal} is not a serious concern in the current application.

\begin{figure}[H]
\centering
\includegraphics[width = \textwidth]{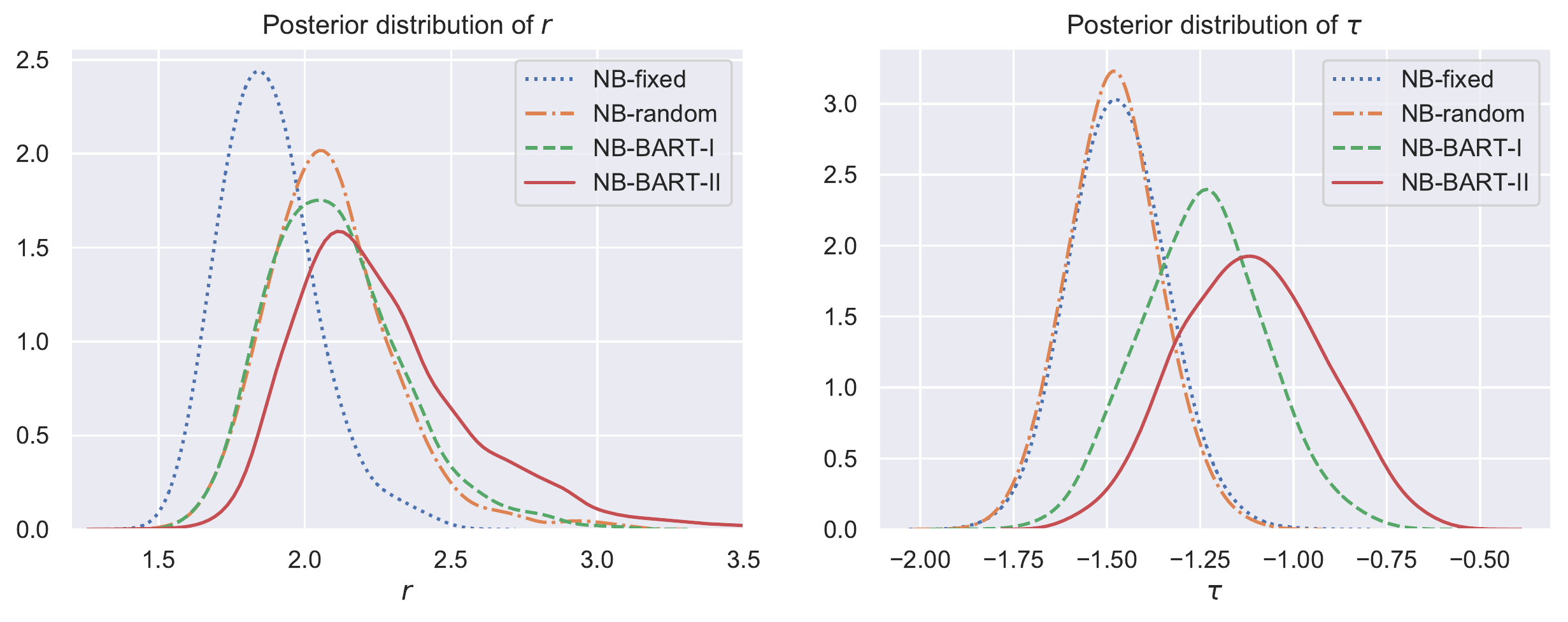}
\caption{Posterior distributions of the negative binomial shape parameter $r$ and the spatial association parameter $\tau$ for year 2010 by model specification.} \label{fig:posterior}
\end{figure}

\begin{table}[H]
\centering
\footnotesize
\input{table_results}
\caption{Estimation results for the negative binomial model with only fixed link function parameters (NB-fixed) and the negative binomial model with both fixed and random link function parameters (NB-random).} \label{tab:results}
\end{table}

\section{Conclusions} \label{section:conclusion}

In this paper, we propose a spatial negative binomial Bayesian additive regression trees (NB-BART) model for the identification of accident hot spots in road networks. The contribution of our work is threefold:
\begin{itemize}
\item First, the proposed model re-conceptualises the specification of the link function in count data models. Since NB-BART specifies the link function as a sum-of-step-functions, it can flexibly account for interactions and non-linear relationships between predictors. This flexibility comes at no expense in interpretability, because a modeller can still specify a semiparametric link function in combination with a BART component and random parameters in a linear component. Furthermore, a modeller does not require to exogenously create dummy variables from continuous covariates before entering them into the link function, because BART endogenously partitions the predictor space and automatically extracts the maximum possible information from the predictors.    
\item Second, we derive a Gibbs sampler for NB-BART by employing the state-of-the-art P{\'o}lya-Gamma data augmentation technique. The sampler ensures conjugacy of the conditional posteriors of all non-BART parameters with the exception of a scalar spatial correlation parameter, which can be updated in a Metropolis-Hastings step. The Bayesian inference approach allows for the construction of credible intervals and other derived quantities such as site rankings.
\item Third, we benchmark the performance of NB-BART against a baseline negative binomial regression model with a linear link function and spatial correlations. Goodness-of-fit and site ranking measures indicate that NB-BART performs as well as or better than the baseline model with random parameters in the linear-in-parameters link function. Our results suggest that if the objectives of a study are centred around prediction, a modeller may be better off considering a flexible link function with a BART component in lieu of a linear-in-parameters link function with random parameters. Nonetheless, if heterogeneity in some parameters is of interest to the modeller, it can be incorporated in NB-BART with a semiparametric link function. At the same time, we acknowledge that it is possible that a count data model with a sophisticated mixing distribution such as a finite mixture-of-normals or Dirichlet process mixture-of-normals mixing distribution \citep[see e.g.][]{buddhavarapu2016modeling, cheng2020exploring, heydari2016bayesian} could effectively control for non-linearities in the predictors and could be more competitive with NB-BART in certain empirical applications. However, such models are time-consuming to estimate and necessitate post-hoc model selection for determining the dimensionality and complexity of the mixing distribution, whereas NB-BART endogenously partitions the predictor space.
\end{itemize}

The hierarchical nature of NB-BART offers opportunities for extensions in two directions. First, a multivariate extension of NB-BART can facilitate the joint modelling of multiple crash frequency variables such as crash counts by crash type, severity level, and vehicle type \citep{dong2014multivariate,yasmin2018multivariate}. Second, the proposed NB-BART formulation accounts for only spatial random effects but can be extended to accommodate temporal variation in model parameters as well as spatiotemporal random effects \citep{li2019hierarchical,liu2017exploring}.    

Recent developments in BART can also be incorporated in the above-discussed extensions. First, in the case of sparse and large predictors spaces, the original BART framework is vulnerable to the curse of dimensionality. Recently proposed soft trees can be adopted in order to handle this challenge \citep{linero2018bayesian}. Second, Bayesian regression trees have recently been used for observational causal inference \citep{hahn2020bayesian}, which is a critical avenue for future research in the accident analysis literature \citep{mannering2020big}.  

\section*{Acknowledgements}
We would like to thank two anonymous reviewers for their critical assessment of our work.


\newpage
\bibliographystyle{apalike}
\bibliography{bibliography.bib}

\end{document}

%% file: table_sample_description.tex
\begin{tabular}{l|ccc|rr|rr|rr|rr}
\toprule
{} & \multicolumn{3}{c|}{\textbf{Predictor space}} & \multicolumn{2}{c|}{\textbf{2007}} & \multicolumn{2}{c|}{\textbf{2008}} & \multicolumn{2}{c|}{\textbf{2009}} & \multicolumn{2}{c}{\textbf{2010}} \\
{} &               \textbf{I} & \textbf{I (random)} &      \textbf{II} &   \textbf{Mean} &  \textbf{Std.} &   \textbf{Mean} &  \textbf{Std.} &   \textbf{Mean} &  \textbf{Std.} &   \textbf{Mean} &  \textbf{Std.} \\
\midrule
Crash count                                        &             N.A. &        N.A. &     N.A. &  19.15 & 25.76 &  15.17 & 20.60 &  14.17 & 19.21 &  17.44 & 23.68 \\
Interstate highway (dummy)                         &          \cmark &     \xmark &  \cmark &   0.45 &  0.50 &   0.45 &  0.50 &   0.45 &  0.50 &   0.45 &  0.50 \\
Exurban area (dummy)                               &          \cmark &     \xmark &  \cmark &   0.27 &  0.45 &   0.27 &  0.45 &   0.27 &  0.45 &   0.20 &  0.40 \\
Asphalt pavement (dummy)                           &          \cmark &     \xmark &  \cmark &   0.17 &  0.38 &   0.17 &  0.37 &   0.14 &  0.35 &   0.12 &  0.33 \\
Asphalt shoulder (dummy)                           &          \cmark &     \xmark &  \cmark &   0.60 &  0.49 &   0.58 &  0.49 &   0.58 &  0.49 &   0.57 &  0.50 \\
Total road width                                   &          \xmark &     \xmark &  \cmark &  54.51 & 15.17 &  55.00 & 15.27 &  55.97 & 15.62 &  56.27 & 15.43 \\
Left shoulder width [ft]                           &          \xmark &     \xmark &  \cmark &   8.61 &  2.78 &   8.66 &  2.78 &   8.36 &  3.33 &   8.52 &  3.23 \\
Right shoulder width [ft]                          &          \xmark &     \xmark &  \cmark &   9.02 &  2.31 &   9.06 &  2.30 &   9.75 &  2.07 &   9.60 &  2.26 \\
Left shoulder width < 10 ft                        &          \cmark &     \cmark &  \xmark &   0.53 &  0.50 &   0.53 &  0.50 &   0.52 &  0.50 &   0.51 &  0.50 \\
Right shoulder width < 10 ft                       &          \cmark &     \cmark &  \xmark &   0.44 &  0.50 &   0.44 &  0.50 &   0.25 &  0.43 &   0.27 &  0.45 \\
Road overall quality index                         &          \xmark &     \xmark &  \cmark &  35.40 & 20.11 &  36.66 & 18.08 &  36.03 & 19.32 &  37.89 & 17.92 \\
Road overall quality index $\leq$ 45               &          \cmark &     \xmark &  \xmark &   0.50 &  0.50 &   0.50 &  0.50 &   0.51 &  0.50 &   0.56 &  0.50 \\
Road comfort index                                 &          \xmark &     \xmark &  \cmark &  34.48 &  5.70 &  34.95 &  5.57 &  34.57 &  5.78 &  35.13 &  5.57 \\
Road structural index                              &          \xmark &     \xmark &  \cmark &  41.80 & 14.95 &  42.52 & 13.71 &  42.69 & 13.97 &  43.56 & 12.76 \\
Road surface index                                 &          \xmark &     \xmark &  \cmark &   0.61 &  1.27 &   0.62 &  1.28 &   1.82 &  1.47 &   1.06 &  1.54 \\
Speed limit [MPH]                                  &          \cmark &     \cmark &  \cmark &  61.17 &  5.05 &  61.25 &  4.92 &  61.35 &  4.85 &  61.37 &  4.79 \\
No. of through lanes                               &          \xmark &     \xmark &  \cmark &   3.13 &  0.99 &   3.16 &  1.01 &   3.15 &  1.03 &   3.18 &  1.01 \\
Road profile score (avg.)                          &          \cmark &     \xmark &  \xmark & 117.45 & 35.76 & 114.40 & 34.47 & 116.89 & 36.10 & 113.11 & 34.19 \\
Road profile score (left)                          &          \xmark &     \xmark &  \cmark & 117.15 & 35.11 & 107.67 & 34.38 & 114.86 & 34.51 & 112.04 & 32.91 \\
Road profile score (right)                         &          \xmark &     \xmark &  \cmark & 117.88 & 37.55 & 121.32 & 37.50 & 119.06 & 38.67 & 114.32 & 36.44 \\
Annual average daily traffic (AADT) [10k veh.] per lane &          \xmark &     \xmark &  \cmark &   1.52 &  0.84 &   1.54 &  0.85 &   1.56 &  0.86 &   1.50 &  0.85 \\
Logarithm of AADT per lane                         &          \cmark &     \xmark &  \xmark &   9.46 &  0.61 &   9.47 &  0.61 &   9.48 &  0.61 &   9.45 &  0.61 \\
Truck traffic percentage                           &          \cmark &     \xmark &  \cmark &  10.49 &  6.73 &  10.36 &  6.47 &  10.69 &  6.57 &  10.79 &  6.36 \\
\bottomrule
\end{tabular}

%% file: table_fit.tex
\begin{tabular}{l |rr|rr|rr|rr}
\toprule
{} & \multicolumn{2}{c|}{\textbf{2007}} & \multicolumn{2}{c|}{\textbf{2008}} & \multicolumn{2}{c|}{\textbf{2009}} & \multicolumn{2}{c}{\textbf{2010}} \\
\textbf{Method} & \textbf{LPPD} &  \textbf{RMSE} & \textbf{LPPD} &  \textbf{RMSE} & \textbf{LPPD} &  \textbf{RMSE} & \textbf{LPPD} &  \textbf{RMSE} \\
\midrule
NB-fixed   &           -3784.34 &           14.74 &           -3545.04 &          11.66 &           -3545.89 &          11.73 &           -3627.46 &           12.80 \\
NB-random  &           -3726.01 &           13.77 &           -3483.54 &          10.85 &           -3478.76 &          10.93 &           -3564.49 &           12.03 \\
NB-BART-I  &           -3734.88 &           13.82 &           -3490.67 &          10.69 &           -3510.13 &          10.97 &           -3558.87 &           11.46 \\
NB-BART-II &  \textbf{-3664.14} &  \textbf{12.15} &  \textbf{-3437.68} &  \textbf{9.84} &  \textbf{-3407.94} &  \textbf{9.39} &  \textbf{-3510.43} &  \textbf{10.53} \\
\midrule
\multicolumn{9}{p{14cm}}{
\footnotesize
Note: LPPD = log pointwise predictive density; RMSE = root mean square error. For each observation period and goodness fit measures, the best-performing method is highlighted in bold font.
} \\
\bottomrule
\end{tabular}

%% file: table_ranking_performance.tex
\begin{tabular}{l | ccc | ccc | ccc}
\toprule
{} & \multicolumn{3}{c|}{\textbf{2007}} & \multicolumn{3}{c|}{\textbf{2008}} & \multicolumn{3}{c}{\textbf{2009}} \\
{} & $\mathcal{T}_{\text{SC}}$ & $\mathcal{T}_{\text{MC}}$ & $\mathcal{T}_{\text{TRD}}$ & $\mathcal{T}_{\text{SC}}$ & $\mathcal{T}_{\text{MC}}$ & $\mathcal{T}_{\text{TRD}}$ & $\mathcal{T}_{\text{SC}}$ & $\mathcal{T}_{\text{MC}}$ & $\mathcal{T}_{\text{TRD}}$ \\
\midrule
EB         &                       28.27 &                 \textbf{41} &                        25.57 &                       28.62 &                          38 &               \textbf{26.34} &                       30.99 &                          27 &                        48.45 \\
NB-fixed   &                       54.95 &                          39 &                        27.03 &                       46.94 &                          39 &                        30.10 &                       60.87 &                 \textbf{31} &                        49.17 \\
NB-random  &                       56.35 &                          40 &               \textbf{23.60} &                       47.48 &                 \textbf{41} &                        29.67 &                       61.10 &                          30 &                        48.41 \\
NB-BART-I  &                       57.52 &                          33 &                        29.83 &                       49.80 &                          40 &                        33.72 &                       61.84 &                          30 &                        59.40 \\
NB-BART-II &              \textbf{61.05} &                          37 &                        38.19 &              \textbf{53.73} &                          40 &                        38.47 &              \textbf{69.09} &                          30 &               \textbf{45.93} \\
\midrule
\multicolumn{10}{p{0.8 \textwidth}}{
\footnotesize
Note: 
$\mathcal{T}_{\text{SC}}$ = site consistency test; $\mathcal{T}_{\text{MC}}$ = method consistency test; $\mathcal{T}_{\text{TRD}}$ = total rank differences test.
For each test and reference period, the best-performing hot spot identification method is highlighted in bold font.
} \\
\bottomrule
\end{tabular}

%% file: table_results.tex
\begin{tabular}{l | cccc | cccc}
\toprule
                                        & \multicolumn{4}{c|}{\textbf{NB-fixed}} & \multicolumn{4}{c}{\textbf{NB-random}} \\
\textbf{Parameter}                                        &                      \textbf{2007} &                      \textbf{2008} &                      \textbf{2009} &                      \textbf{2010} &                      \textbf{2007} &                      \textbf{2008} &                      \textbf{2009} &                      \textbf{2010} \\
\midrule
Intercept &   2.00\textsuperscript{*} &   1.70\textsuperscript{*} &   1.69\textsuperscript{*} &   1.72\textsuperscript{*} &   1.90\textsuperscript{*} &   1.59\textsuperscript{*} &   1.58\textsuperscript{*} &   1.61\textsuperscript{*} \\
Interstate highway &   0.44\textsuperscript{*} &   0.42\textsuperscript{*} &   0.35\textsuperscript{*} &                      0.30 &   0.44\textsuperscript{*} &   0.40\textsuperscript{*} &   0.34\textsuperscript{*} &                      0.29 \\
Asphalt pavement &                      0.07 &                      0.08 &                      0.12 &                      0.10 &                      0.06 &                      0.09 &                      0.11 &                      0.09 \\
Exurban area &  -1.03\textsuperscript{*} &  -0.98\textsuperscript{*} &  -0.87\textsuperscript{*} &  -0.83\textsuperscript{*} &  -1.03\textsuperscript{*} &  -0.96\textsuperscript{*} &  -0.86\textsuperscript{*} &  -0.83\textsuperscript{*} \\
Asphalt shoulder &                      0.02 &                      0.10 &                      0.04 &                      0.01 &                      0.03 &                      0.11 &                      0.03 &                      0.01 \\
Road overall quality index $\leq$ 45 &                      0.16 &                      0.07 &                      0.19 &                      0.13 &                      0.15 &                      0.06 &                      0.18 &                      0.13 \\
Truck traffic percentage &                     -0.16 &  -0.26\textsuperscript{*} &  -0.25\textsuperscript{*} &  -0.19\textsuperscript{*} &   0.42\textsuperscript{*} &   0.40\textsuperscript{*} &   0.44\textsuperscript{*} &   0.53\textsuperscript{*} \\
Road profile score (avg.) &   0.15\textsuperscript{*} &   0.12\textsuperscript{*} &                      0.11 &   0.24\textsuperscript{*} &                     -0.16 &  -0.26\textsuperscript{*} &  -0.25\textsuperscript{*} &                     -0.18 \\
Log of AADT per lane &   0.41\textsuperscript{*} &   0.39\textsuperscript{*} &   0.43\textsuperscript{*} &   0.52\textsuperscript{*} &   0.13\textsuperscript{*} &                      0.10 &                      0.10 &   0.24\textsuperscript{*} \\
Speed limit  &                           &                           &                           &                           & \\ 
\quad Loc. &                     -0.06 &                     -0.05 &                     -0.02 &                      0.01 &                     -0.07 &                     -0.05 &                     -0.01 &                      0.00 \\
\quad Scale &                           &                           &                           &                           &   0.11\textsuperscript{*} &   0.10\textsuperscript{*} &   0.15\textsuperscript{*} &   0.15\textsuperscript{*} \\
Left shoulder width < 10 ft &                           &                           &                           &                           & \\ 
\quad Loc. &                     -0.05 &                     -0.13 &                     -0.16 &  -0.40\textsuperscript{*} &                     -0.05 &                     -0.11 &                     -0.20 &  -0.41\textsuperscript{*} \\
\quad Scale &                           &                           &                           &                           &   0.26\textsuperscript{*} &   0.25\textsuperscript{*} &   0.26\textsuperscript{*} &   0.20\textsuperscript{*} \\
Right shoulder width < 10 ft &                           &                           &                           &                           & \\
\quad Loc. &                     -0.31 &                     -0.26 &  -0.28\textsuperscript{*} &                     -0.10 &  -0.34\textsuperscript{*} &  -0.37\textsuperscript{*} &  -0.37\textsuperscript{*} &                     -0.14 \\
\quad Scale &                           &                           &                           &                           &   0.20\textsuperscript{*} &   0.32\textsuperscript{*} &   0.43\textsuperscript{*} &   0.29\textsuperscript{*} \\
Negative binomial shape parameter $r$ &    1.47\textsuperscript{*} &   1.64\textsuperscript{*} &   1.58\textsuperscript{*} &   1.91\textsuperscript{*} &   1.60\textsuperscript{*} &   1.82\textsuperscript{*} &   1.79\textsuperscript{*} &   2.12\textsuperscript{*} \\
Spatial association parameter $\tau$  &  -1.50\textsuperscript{*} &  -1.43\textsuperscript{*} &  -1.42\textsuperscript{*} &  -1.45\textsuperscript{*} &  -1.52\textsuperscript{*} &  -1.47\textsuperscript{*} &  -1.47\textsuperscript{*} &  -1.47\textsuperscript{*} \\
\midrule
\multicolumn{9}{l}{
\footnotesize
\textsuperscript{*} At least 95\% of the posterior mass exclude zero.
} \\
\bottomrule
\end{tabular}